\documentclass[sigconf]{acmart}

\usepackage{comment}
\usepackage{csquotes}
\usepackage{tabularx}
\usepackage[htt]{hyphenat}
\usepackage{xcolor}
\usepackage{graphicx}
\usepackage{hyperref}
\usepackage{subcaption}

\newcommand{\eg}{\textit{e.g.}}
\newcommand{\ie}{\textit{i.e.}}

\newcommand{\etal}{\textit{et al.}}

\newcommand{\leftcell}[2][l]{%
  \begin{tabular}[#1]{@{}l@{}}#2\end{tabular}}

\newlength\myindent
\AtBeginDocument{%
  \providecommand\BibTeX{{%
    \normalfont B\kern-0.5em{\scshape i\kern-0.25em b}\kern-0.8em\TeX}}}

\copyrightyear{2024}
\acmYear{2024}
\setcopyright{rightsretained}
\acmConference[CHI '24]{Proceedings of the CHI Conference on Human Factors in Computing Systems}{May 11--16, 2024}{Honolulu, HI, USA}
\acmBooktitle{Proceedings of the CHI Conference on Human Factors in Computing Systems (CHI '24), May 11--16, 2024, Honolulu, HI, USA}
\acmDOI{10.1145/3613904.3642266}
\acmISBN{979-8-4007-0330-0/24/05}

\begin{document}

\title[From Paper to Card]{From Paper to Card: Transforming Design Implications with Generative AI}

\author{Donghoon Shin}
\orcid{0000-0001-9689-7841}
\email{dhoon@uw.edu}
\affiliation{%
  \institution{University of Washington}
  \city{Seattle}
  \state{WA}
  \country{USA}}

\author{Lucy Lu Wang}
\orcid{0000-0001-8752-6635}
\email{lucylw@uw.edu}
\affiliation{%
  \institution{University of Washington,\\Allen Institute for AI}
  \city{Seattle}
  \state{WA}
  \country{USA}}

\author{Gary Hsieh}
\orcid{0000-0002-9460-2568}
\email{garyhs@uw.edu}
\affiliation{%
  \institution{University of Washington}
  \city{Seattle}
  \state{WA}
  \country{USA}}

\renewcommand{\shortauthors}{Shin et al.}

\begin{abstract}
Communicating design implications is common within the HCI community when publishing academic papers, yet these papers are rarely read and used by designers. One solution is to use design cards as a form of translational resource that communicates valuable insights from papers in a more digestible and accessible format to assist in design processes. However, creating design cards can be time-consuming, and authors may lack the resources/know-how to produce cards. Through an iterative design process, we built a system that helps create design cards from academic papers using an LLM and text-to-image model. Our evaluation with designers ($N=21$) and authors of selected papers ($N=12$) revealed that designers perceived the design implications from our design cards as more inspiring and generative, compared to reading original paper texts, and the authors viewed our system as an effective way of communicating their design implications. We also propose future enhancements for AI-generated design cards.
\end{abstract}



\maketitle

\section{Introduction}

One of the key goals for research in an applied field such as human-computer interaction (HCI) is to generate insights that can directly impact practice---in our case, design practice. Many authors attempt to achieve this through the \textit{design implication} section in an academic paper. However, despite best efforts in communicating these insights, studies have found that design implications, as presented in academic papers, are rarely consumed and used by design practitioners~\cite{colusso2017translational}. Designers describe difficulties reading the often jargon-filled papers and not finding academic papers useful~\cite{colusso2017translational}.

This has led to several calls to develop better translational resources to bridge the gap between research and practice~\cite{colusso2017translational, colusso2019translational}. Instead of relying on academic papers to communicate valuable research findings to design practitioners, what may be needed are other types of resources to support this translation. One approach is to distill the insights into \textit{design cards}. Design cards are tools that facilitate the design process, and can contain several types of design knowledge---from human insights to material \& domain knowledge~\cite{hsieh2023designcards}. There is a long tradition of creating design cards to communicate research insights~\cite{alkhuzai2021evaluating, chung2015understanding}. 

However, while design cards may be more accessible and provide more prescriptive insights to practitioners, one challenge with relying on using design cards to communicate academic findings is how these cards should be generated. It is unrealistic to expect researchers to do it, as HCI researchers describe the lack of time and competing responsibilities that prevent them from doing more outreach work~\cite{williams2022hci}. They also lack the proper incentive to do this type of translational work~\cite{colusso2019translational}. Further, even if researchers are motivated, they may have limited understanding of what designers are seeking in design implications, and lack the skills needed to generate these cards so that they are inspiring~\cite{sas2014generating}. A potential solution may be to leverage crowdwork---ideally crowds of designers---to perform this translational work; this approach has been explored for the creation of videos for research talks~\cite{vaish2018creating}. Nonetheless, changes in incentives are required for long-term sustainability. Furthermore, reliance on volunteer contributors can lead to spotty coverage and a biased focus on certain topics over others~\cite{denning2005wikipedia}.  

In this paper, we propose the use of generative AI models to support this important translational work. Generative AI models, such as large language models (LLMs) and text-to-image models, have recently shown remarkable capabilities in producing high-fidelity texts and images, thus being adapted for various creative tasks~\cite{chung2022talebrush,shakeri2021saga,ko2022large}. Motivated by this line of research, we believe generative AI models could provide a scalable and efficient way of translating academic findings into a prescriptive format such as design cards, without requiring significant time or effort from researchers and designers.

\begin{figure*}[t!]
    \centering
    \includegraphics[width=.625\textwidth]{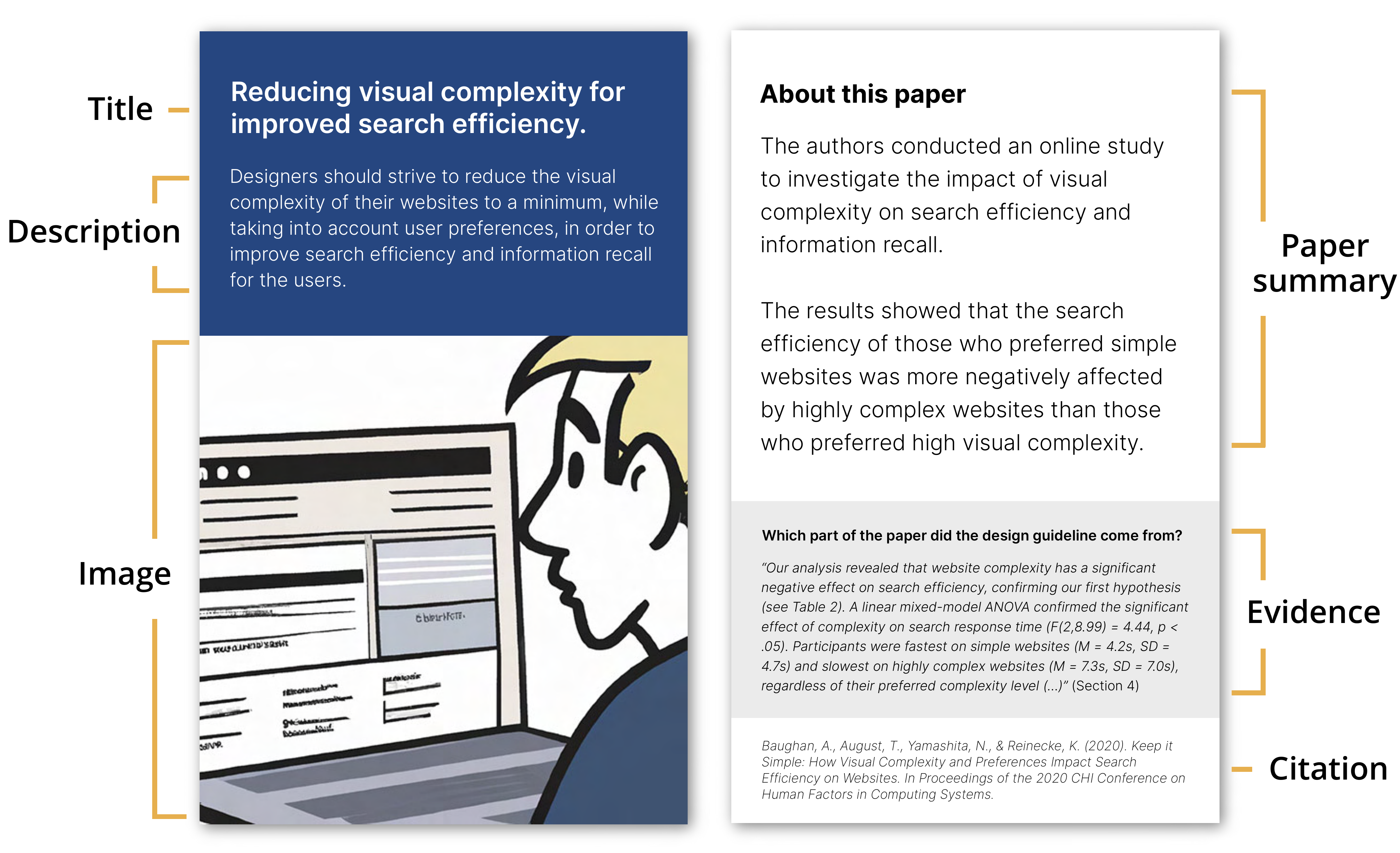}
    \caption{A design card built on Baughan \etal{}~(2020)~\cite{baughan2020keep} using our system}
    \label{fig:example}
    \Description{This figure illustrates a design card built on Baughan et al. (2020) using our system, which is double-sided. This figure illustrates the front side of the card on the left side, which consists of the title, description, and image. On the right side, the figure illustrates the back side of the card, which consists of a paper summary, evidence, and citation.}
\end{figure*}

To test our hypothesis, we develop and evaluate an end-to-end system to automatically generate design cards from HCI publications. We first ran a preliminary interview study with design practitioners ($N=12$) to inform the design of design cards built on academic papers. We presented participants with mock-ups of design cards that we created from three HCI papers, each with several design elements and alternate formats. Based on our findings, we iterated on our design and built a system that automatically generates a design card from the design implication text extracted from a paper using LLMs (\ie{}, GPT-3~\cite{brown2020language}) and text-to-image models (\ie{}, DALL-E~2~\cite{ramesh2022hierarchical}).

To test our system and explore future enhancements of AI-generated design cards, we evaluated a set of generated cards through a survey study with designers ($N=21$) and the authors of HCI papers ($N=12$). We found that designers perceived content from AI-generated design cards as more inspiring and generative compared to the text of design implications sections, and did not consider the generated content to be different in validity, generalizability, or originality. The paper authors further reported that the contents in a design card were overall clear and accurate, and highlighted the potential of cards in strengthening the delivery of their contributions in HCI. In addition to this positive feedback, conversations with designers and authors also surfaced future directions for enhancing AI-generated design cards.

Our work offers several important contributions:

\begin{itemize}
    \item Surfacing design elements of what design practitioners want from design cards as a form of translational resource
    \item Implementing a system that helps streamline the generation of design cards from HCI papers using generative AI models
    \item Survey results with designers and HCI paper authors demonstrating the potential added inspirability and generativity of design cards generated by LLMs, as well as the potential of our system as a delivery tool for authors to present their contributions
    \item Insights for the future enhancement of AI-generated design cards
\end{itemize}
\section{Related Work}

Generative AI models are powerful tools that have been incorporated into many creative and practical tasks. Large language models (LLMs) such as GPT are trained on large amounts of unlabeled text data and human feedback, and are capable of generating highly consistent and contextually relevant text across a variety of domains and tasks, such as robot journalism~\cite{pavlik2023collaborating}, customer service~\cite{shrivastava_2023}, and healthcare~\cite{sezgin2022operationalizing}. Moreover, the generative capabilities of LLMs have been featured as a potential way to inspire and support creative, in addition to practical, tasks. For example, prior studies have explored the use of LLMs in supporting creative writing~\cite{chung2022talebrush, calderwood2020novelists}, idea generation~\cite{di2022idea}, and generating song lyrics~\cite{watanabe2018melody}.

Text-to-image models are another type of generative AI model that has been widely used for a range of applications~\cite{gal2022image}. Using deep learning techniques to generate or modify images, text-to-image models allow users to create various illustrations, from realistic depictions of landscapes to abstract art. A well-known example of a text-to-image model is DALL-E 2~\cite{ramesh2022hierarchical}, a transformer-based diffusion model that produces images based on descriptive textual prompts. Other latent diffusion models such as Stable Diffusion~\cite{rombach2021highresolution} and Midjourney\footnote{https://www.midjourney.com} have also been shown to effectively generate images based on textual prompts. These models have been applied to various pragmatic and creative application areas, such as fine-art painting (\eg{}, r/ImagenAI\footnote{https://www.reddit.com/r/ImagenAI}, r/aiArt\footnote{https://www.reddit.com/r/aiArt}), fashion design~\cite{ak2019semantically}, and webtoon sketching~\cite{ko2022we}.

Generative AI models have been used to make scientific literature more accessible and efficient to consume. As generative AI models are capable of producing contextually relevant information (\eg{}, summaries, visual representations), researchers have highlighted their use in transforming or augmenting the reading of scholarly articles to allow a wider audience to gain knowledge from research. For instance, previous studies have demonstrated the use of generative AI models for facilitating paper consumption, on tasks such as paper summarization~\cite{cachola-etal-2020-tldr, li-etal-2020-cist, deyoung-etal-2021-ms, lu-etal-2020-multi-xscience}, generation of research highlights~\cite{rehman-etal-2022-named}, and automatic generation of posters or slides~\cite{Sefid2019AutomaticSG, Fu2021DOC2PPTAP, Xu2022PosterBotAS}. Not limited to these approaches of supporting general paper-reading practices, generative AI models have also been shown to be effective in augmenting domain-specific papers. For example, August \etal{}~proposed a tool called Paper Plain~\cite{august2022paper}, which leverages an LLM to lower the barrier to consuming medical knowledge by augmenting medical papers with features such as section summaries and passage-specific question-answering.

Motivated by this research, we highlight in this study the potential of generative AI models in communicating design implications presented in HCI papers. Sas \etal{}~\cite{sas2014generating} suggested several evaluation criteria for assessing the quality of design implications: \textit{theoretical/empirical validity} (\ie{}, the level of being academically grounded), \textit{generalizability} (\ie{}, applicability beyond the fieldwork), \textit{originality} (\ie{}, ensuring that the design implication is original), \textit{generativity} (\ie{}, ability to open up new design spaces), \textit{inspirability} (\ie{}, ability to stimulate designers to explore further), and \textit{actionability} (\ie{}, ability to be acted upon); that is, communicating design implications requires the conveyance of both creativity and academic value. As previous literature has shown the feasibility of generative AI models in generating both contextually relevant and creative outputs, we believe that these characteristics will help augment the utility (\ie{}, generativity, inspirability, actionability) of the academic paper without losing its scholarly robustness (\ie{}, validity, generalizability, originality)~\cite{chi2023} when generating design cards, thus assisting designers to more easily consume design insights from academic papers.
\section{Preliminary Study}

To understand what designers want to see from design cards for gaining design insights from academic papers and guiding design, we first conducted an interview study with 12 designers.

Although there are no fixed rules for designing design cards, several common components have been used in previous design cards~\cite{aarts2020design}, presumably due to the limited size of design cards, their physical format, and the emphasis on the visibility of each component. Thus, we began by designing an initial set of design cards that contain several frequently occurring components to guide participants in offering feedback.

\subsection{Initial Design of Design Cards}

To choose papers for our initial design cards, we searched in the ACM Digital Library using keywords related to design implications (\eg{}, design implication, design guideline, future design). From the results, we filtered for full-paper publications from ACM-sponsored conferences/journals that include design implication sections. Among these, we chose three papers with non-overlapping topics that offer implications across a range of design practices (\ie{}, automotive UX design~\cite{schneider2021explain}, online security~\&~safety for children~\cite{mcnally2018co}, and mobile UI design~\cite{chaudry2012mobile}).

First, we performed a cursory exploration of the capabilities of generative models (\ie{}, summarization, image generation, question answering, instruction-based text completion, in-context learning) as well as rule-based techniques (\eg{}, deriving a bibliographic reference using a paper's DOI) to determine if each component of the design card can be generated using these capabilities. We then prepared a set of initial design cards consisting of potentially feasible components that are commonly found in existing design cards~\cite{ideo_methodcards,lockton2010design,casais2016using,deng2014tango, yoon2016developing, colusso2018behavior}. Our initial design cards included a title, description, image, source text of the design implication, paper summary (concise overview of the whole paper), and citation. At this stage, each component was manually generated and prepared by the researchers except for images, which were created using a text-to-image model (DALL-E 2). For some components, we offered multiple options to let participants explore diverging styles (see examples in \autoref{fig:prelimcard}). Descriptions for each component are as follows:

\begin{figure*}
    \centering
    \includegraphics[width=\textwidth]{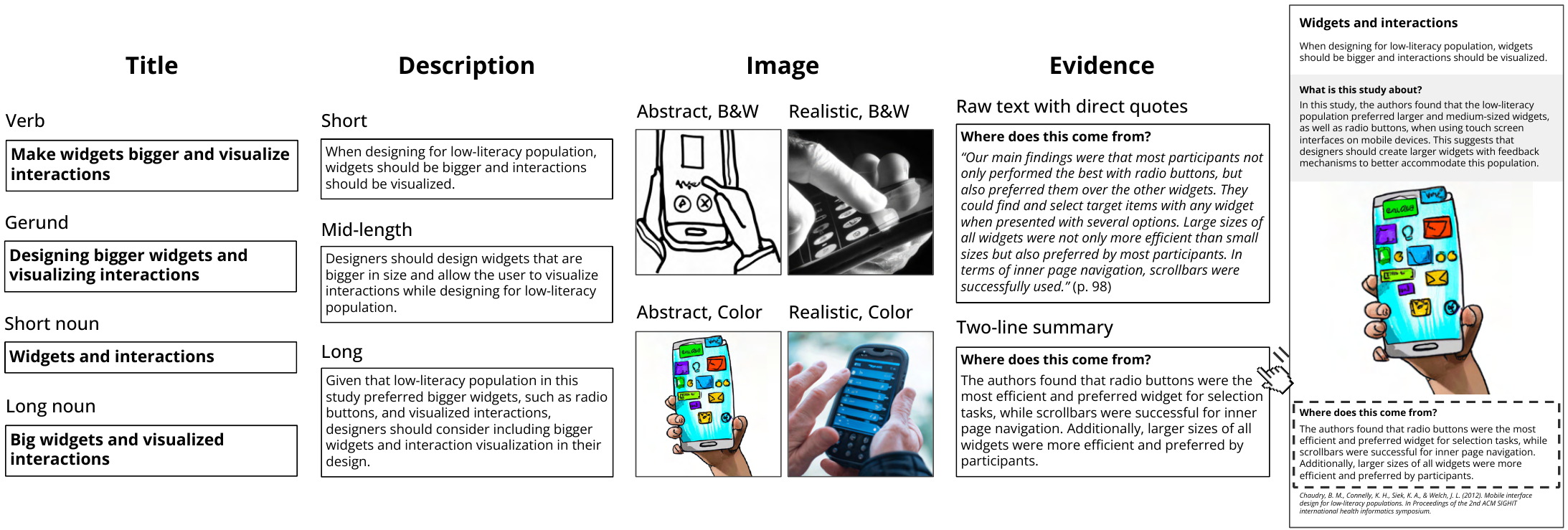}
    \caption{An illustration of the Miro board participants used during the preliminary study. Each participant was asked to choose one among several options for each component, and drag \& drop it to a card template to make a design card.}
    \label{fig:prelimcard}
    \Description{This figure illustrates the Miro board that participants used during our preliminary study. This board contains options for title, description, image, and evidence. The title has four options, which are (i) a verb-based title, (ii) a gerund-based title that starts with a gerund, (iii) a short noun-style title, and (i) a long and more descriptive noun-style title. Similarly, the description has three options: (i) short, (ii) mid-length, and (iii) long description. The image had four options: (i) abstract and black \& white, (ii) realistic and black \& white, (iii) abstract and colored, and (iv) realistic and colored. Lastly, the evidence had two options: (i) a raw evidence text with direct quotes and (ii) a two-line summary of the evidence. Participants can drag and drop each component into the card template to make a design card.}
\end{figure*}

\paragraph{Title}

The title of the design card is meant to provide an easy-to-understand, high-level description of the design implication. To understand the designer's preference around the title's actionability, length, and grammatical format, we set up four types of titles for each paper: verb-format, gerund-format, long-noun, and short-noun. Verb- and gerund-format were designed to describe an action of a design guideline (\eg{}, \texttt{Make widgets bigger and visualize interactions} / \texttt{Designing bigger widgets and visualizing interactions}), while the goal of long- and short-noun was mainly to convey the design target and context in a keyword (\eg{}, \texttt{Widgets and interactions} / \texttt{Big widgets and visualized interactions}).

\paragraph{Description}

The description of the design card aims to offer detailed guidance on following the design implication. To probe the details that designers want to see in the description of a design implication, we decided to let designers explore the formats of this component by preparing three options for the description: \texttt{short} (approx. 10-15 words), \texttt{mid-length} (approx. 20 words), and \texttt{long} (approx. 30 words) descriptions, each of which varies in the level of detail.

\paragraph{Image}

Image is a visual illustration that is included to inspire and to help designers gain a sense of the context of a design implication. Prior literature suggested that the use of color in images has been shown to capture the attention and interest of viewers, yet it can also be distracting and hinder comprehension of information~\cite{lamberski1983instructional}. Similarly, although photorealistic images offer a higher level of believability, abstracting an image is frequently employed to effectively communicate image cues and enhance the viewer's perception of the image's subject~\cite{haeberli1990paint}. To test these two dimensions in design cards, we prepared images by varying two stylistic dimensions: \textit{abstractness} and \textit{color}.
 
Specifically, we decided to use a text-to-image model instead of retrieving from existing image repositories, because academic papers often introduce new concepts and use cases for which there is no suitable existing illustration. We set up combinations of two specifications for each dimension, resulting in four styles: \texttt{abstract colored}, \texttt{abstract black \& white (B\&W)}, \texttt{realistic colored}, and \texttt{realistic B\&W}. Using several prompts involving the paper title and specified style, we generated images for each option in our study using DALL-E 2~\cite{ramesh2022hierarchical}.

\paragraph{Paper summary}

As each design implication is rooted in the research context, we believe it is important to present the study context in the design card. Thus, we summarized the abstract of the source paper in two sentences and included it in our design. Unlike other components, we did not present multiple options for the paper summary.

\paragraph{Evidence of the design implication}

The purpose of evidence is to let readers review the foundation that underpins the design guideline. From a part of the paper that contains the result of the study, we identified a paragraph providing evidence that supports the design guideline. Particularly, we wanted to determine whether designers would feel the design card to be more grounded when given the original evidence text from the paper versus a summary of the paragraph written in plain language. Thus, we prepared two options that can potentially be used as evidence: \texttt{raw text of the evidence with direct quotes} and a \texttt{summary of the evidence written in plain language}.

\paragraph{Citation}

In cases where designers want to more closely examine the paper, we included a reference to the paper at the bottom of the design card. This could be designed as a clickable link to the paper if the card is presented in digital format. Similar to the paper summary, we also provided only one version of the citation in APA format.

\subsection{Recruitment \& Participants}

To recruit designers, we posted our study recruitment in two design-focused online communities at our university. We recruited 12 participants (\autoref{tab:participants}): 7 of the participants had more than 3 years of prior design work experience, 3 participants had 1--3 years of experience, and 2 participants had no prior work experience. Of these participants, 7 were working at a company as a designer, and 5 were pursuing a design degree at the time of the study. The average age of participants was 25.3 ($SD=3.4$); 9 self-identified as female, 2 non-binary, and 1 male.

\subsection{Study Procedure \& Analysis}

We prepared a Miro board (a collaborative diagramming tool) for each paper option, which contains (i) the options for each component and (ii) a card template where participants can drag \& drop options to build a design card.

Studies were conducted remotely on Zoom. We first consented participants to the study and then briefly explained the goal of the study and the design cards to each participant. Using the Zoom chat functionality, we asked participants to choose one among the three paper titles used in this study that is most relevant to their design interests. We sent participants the paragraph of the paper that contains the design implication and asked them to read it.

Following, we sent participants a web link that directed them to the Miro board of the paper they chose. Once they were on the Miro board, for each design card component, participants were asked to drag \& drop one among several options they prefer the most to the card template, and explain their reason for choosing it (\autoref{fig:prelimcard}). We then asked them to evaluate the design card based on metrics for evaluating design implications~\cite{sas2014generating} (\ie{}, empirical / theoretical validity, generalizability, originality, generativity, inspirability, and actionability), and discuss any suggestions for enhancing the design cards based on each metric. Subsequently, we showed participants the design cards for the other two papers with the same configurations they chose, and asked them to evaluate the other cards as well, to help enrich our evaluation.

Each interview lasted approximately 45 minutes. Once complete, each participant was compensated 20 USD for their participation. All the procedures of our study were reviewed and approved by the IRB of our university's human subjects division.

To analyze the data, we first transcribed the voice recordings. Then, we analyzed the results by using thematic analysis~\cite{braun2006using}, where the researchers first identified and categorized the significant themes from the responses. Then, we discussed the initial themes and refined them until we reached a consensus on the final themes.

\subsection{Results}

We first outline the themes related to the overall design of design cards. Following that, we report the themes categorized by each design card component.

\subsubsection{Overall structure and design of design cards}

Overall, participants were satisfied with the types and structure of information in our design card: \textit{“I like the structure (of design cards). I like how title, description, and all the components make sense with each other.”}~(P7) At the same time, we identified potential improvements around the visual representation of the design card:

\paragraph{Distributing textual information to avoid information overload}

First, displaying every component on a single side of the design card made it difficult for participants to read. As our initial design card template puts all content on a single page, this was reported to affect the visibility of each component, ultimately making the readers overwhelmed and hesitant to dig deeper: \textit{“(On a single page) there are a lot of texts, and that's something that I am overwhelmed with before reading (each component).”} (P3)

To resolve this, we explored the potential of using the back side of design cards to distribute the textual information. Participants mentioned that it would make the cards less overwhelming to put important components on the front and the rest on the back side: \textit{“Let's say that we have another side of this card, and we can include some stuffs in another side of the design card.”} (P8)

\paragraph{Importance of visual aesthetics for augmenting inspirability}

In addition, participants described the importance of visual aesthetics (\eg{}, vivid background color) for inspiring designers. They reported that the visually pleasing design of cards is a major factor that induces them to take a closer look at contents and think through them: \textit{“I want the cards to be more visually attractive and it will make me feel creative and inspired (...) I once saw very pretty design cards and they made me want to use them.”} (P2)

\subsubsection{Title}

\paragraph{Title as a means of guiding action without being directive}

As the title is the first component that designers read while glancing at the design cards, participants spoke to the importance of the title in delivering actionable insights in a concise way. In other words, instead of simply offering keywords (nouns) related to the design context or subject, they preferred to read which action they should take from the title. For these reasons, 10 out of 12 participants preferred either gerund or verb format, as these grammatical constructs are perceived as actions: \textit{“Seems like they (verb and gerund) are telling me what I'm supposed to do so they are clearer.”}~(P6)

At the same time, participants mentioned that titles starting with a verb are too directive, potentially constraining the scope and generalizability of the use of design cards in their future designs. Instead, gerund format was mentioned as a way of delivering guidance to act, without forcing the readers to follow the action: \textit{“Verb seems like it's directing me to must follow a certain thing like as expecting (...) so I think gerund format would make more sense.”}~(P5)

\subsubsection{Description}

Designers mainly viewed the role of description as a `tool for understanding the title more in a detailed way.' Specifically, we elicited from designers several necessary elements of a description that they wished to see:

\paragraph{Desire to know about the target users}

Participants reported that the target user of the design implication (\eg{}, low-literacy population, drivers, children) should be clearly noted in the description, unless it is general users. This is reported to help them identify the scope and limitation of the design implication, and determine if the design implication is applicable to their current designing target: \textit{“The target user should be included because I would want to know who I'm doing it for.”}~(P10)

\paragraph{Identifying the context (\ie{}, design subject) in which the design implication can be applied}

Similarly, design subject (\ie{}, specific object or system that the design implication can be applied to) was reported as an essential component that should be clearly indicated in the design implication description. It could be a physical object (\eg{}, widget, ride), or an intangible system (\eg{}, monitoring systems): \textit{“(I prefer mid-length because) it has a context of which system designers are designing.”}~(P7)

\paragraph{Justification for following the design implication}

Lastly, they would like to see ‘why’ they should follow a certain design implication. Even though citations make the design implication believable, having a reasoning of why the design should be a certain way (\eg{}, \textit{Designers should AAA, \textbf{in order to BBB}}) helps them perceive the implication as valid and gain more confidence in following the recommendation: \textit{“(in a short description) There is no justification for why you should follow it (...) I won't believe in following the guideline without justification.”}~(P2)

\subsubsection{Image}

\paragraph{Representing a potential usage context}

As an image is the only visual illustration among the components, participants mentioned the importance of images in imagining the overall context of the design implication. Thus, they wanted to see how the potential users might be using the design where the design implication can be applied: \textit{“If someone's not immediately reading all of the text, just looking at the image should give them a context or like they're able to imagine a picture themselves in this scenario where the design implication is applied.”}~(P10)

\paragraph{Color to inspire creativity, but without distorted representation}

Except for one participant, everyone responded that they preferred colored illustrations for images, by speaking out to the importance of using color in an image to stimulate inspirability: \textit{“It (colored image) makes it visually appealing (...) the most amount of my attention is very clear when it's in color in some way.”}~(P6) However, participants also expressed concern about the potential for distorted elements in realistic images after noticing distorted representations of human faces, which is a known issue in AI-generated images~\cite{Heusel2017GANsTB, Borji2022GeneratedFI}. Therefore, they preferred colored images for the inspirational value, but more abstract as to not compromise image authenticity: \textit{“I noticed those faces (distorted).. maybe an abstract one is better if you can't do realistic.”}~(P3)

\subsubsection{Paper summary, evidence of the design implication, and citation}

Overall, participants found the paper summary and citation presented in the design card to be necessary. First, a summary of the paper was reported to help them better understand the overall context of the paper that the design implication is built on: \textit{“I really like the paper summary because I can actually see what the study is about.”}~(P4) Also, the citation helped them perceive that the design implication was from a believable source: \textit{“(By seeing the citation) I feel like it's grounded and build trust on the contents.”}~(P6)

For the evidence of design implication, a majority of the participants (8 out of 12) preferred the raw text with direct quotes instead of a summary in plain language, due to the credibility of the format: \textit{“It (direct quotes for the evidence) makes the design implication more formal and credible.”}~(P1)
\section{Design Iteration \& Implementation}

\begin{figure*}
    \centering
    \includegraphics[width=\textwidth]{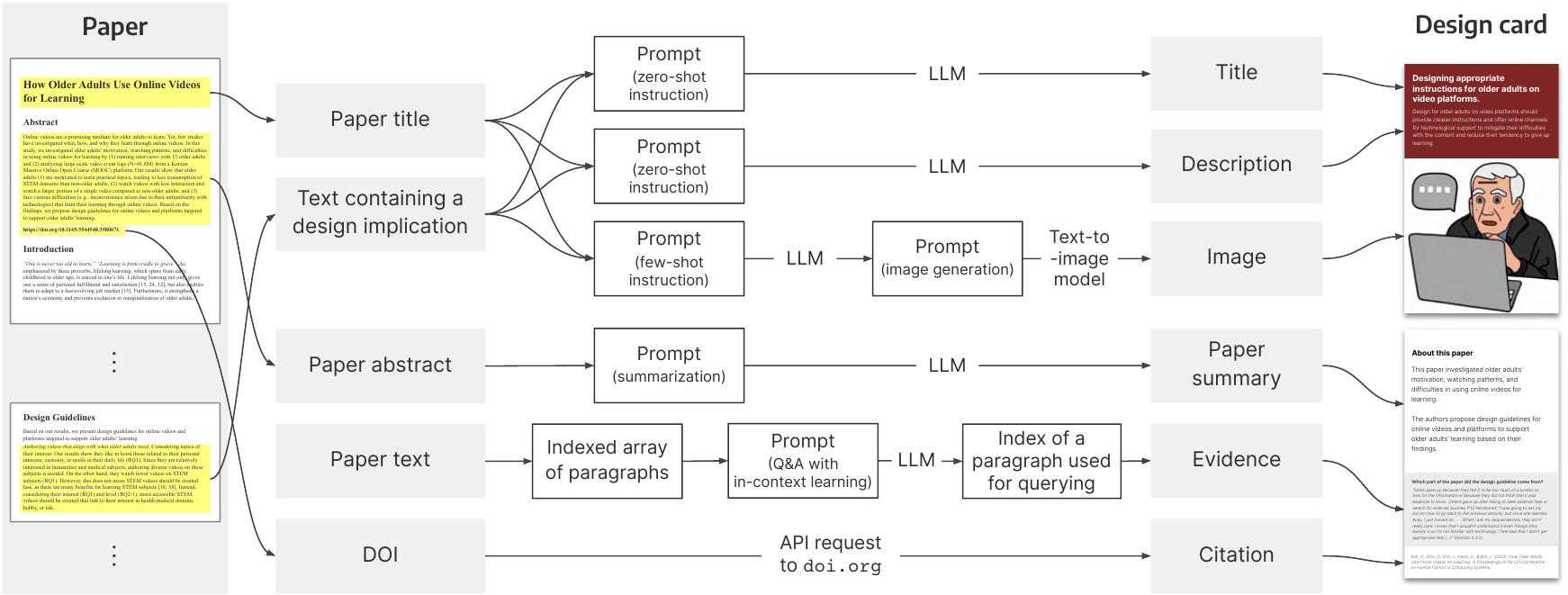}
    \caption{Overview of the pipeline for generating design cards using generative AI models. In this case, we use GPT-3 as the LLM and DALLE-2 as the text-to-image model.}
    \label{fig:implementation}
    \Description{This figure illustrates the pipeline for generating design cards using generative AI models. The title and the description are generated by feeding the original paper's title and the design implication text to an LLM using a zero-shot instruction. The image is generated by first feeding the original paper's title and the design implication text to an LLM using a few-shot instruction to generate an image prompt, which is then fed into a text-to-image model. The paper summary is generated by feeding the paper abstract into the LLM and prompting the model to summarize it. The evidence is generated by first indexing the paper text into an array of paragraphs, and running the Q\&A task with in-context learning using LLM to get the index of a paragraph used for querying. Lastly, the citation is generated by sending the API request to doi.org API with the DOI as a parameter.}
\end{figure*}

From the preliminary study, we uncovered insights on how to better structure and design the cards to support designers' understanding and potential use of the design implications. Based on these results, we iterated on the design card layout and content. We also implemented a pipeline to automatically generate design cards from academic papers, which we discuss below. An example design card generated by our system is illustrated in \autoref{fig:example}.

\subsection{Overall Structure and Design}

First, to alleviate information overload stemming from displaying all information on a single side of a card, we extended our design to use both sides of a card. Specifically, we put core design card components (\ie{}, title, description, image) on the front side, and all the other components (\ie{}, paper summary, evidence, citation) on the backside.

In addition, to improve upon aesthetics and achieve consistent styling, we programmed our system to match the background color of the front side with the most dominant color of the generated image. Specifically, our system obtains a dominant color from the generated image using the modified mean cut quantizer technique~\cite{bloomberg2008color}, adjusts the color to a high level of saturation and low level of value in the HSV color dimension, and sets this as the background color. This results in a vivid and visually engaging front side with a clearly readable card title and description.

\subsection{Generating Card Components} \label{sec:generating_components}

\subsubsection{Title}

Given participants' input, we focused on making the title actionable. As such, we decided to generate the title using an LLM with an instructional prompt specifying that the output include an action verb in the form of a gerund phrase.

While engineering the prompt, we noticed that some paper authors included the target context of their design (\eg{}, user, design subject) in the paper title while omitting them in their design implication text. Simply relying on the design guideline text may omit salient keywords from the paper, so we also included the paper title in the LLM prompt to encourage centering of the output around the keywords in the title (Appendix~\ref{appendix:prompt:title}). To verify that the output is formatted as intended, we programmatically detect that the starting word of the output is in gerund format.

\subsubsection{Description}

For the description, we focused on surfacing the target user, design subject, and justification for following the design implication. Thus, similar to the title, we set up a prompt instruction for an LLM to generate a one-line description for designers such that the output includes these elements (Appendix~\ref{appendix:prompt:description:text}). To check if the output satisfies the requirements we included in the prompt instruction, we used the LLM to evaluate its own output (inspired by work such as~\cite{wu2023large, bai2022constitutional, liu2023gpteval}). Specifically, we programmed our system to verify the results as another classification task using an LLM. For example, to ensure that the justification for design is included in the output description, the system prompts the model to verify as shown in  Appendix~\ref{appendix:prompt:description:verification}. If one of our three requirements are not met, we prompt the system to generate three additional outputs and select from among them the best output based on the same criteria.

\subsubsection{Image}

To generate an image, we used both an LLM and a text-to-image model. Specifically, (i) we first prompt the LLM to generate an image prompt that illustrates the context of the design implication; (ii) we then use the text-to-image model with the prompt output from (i) to generate an image.

For implementation, we engineered a few-shot prompt that contains three training pairs of source design implication text and the corresponding text-to-image prompt describing its usage context~\cite{riche2010studying, kim2023older, kulp2019comparing}. We noticed, through our prompt engineering explorations, that jargon terms (\eg{}, MOOC; massive open online course) were often difficult for text-to-image models to interpret and were more likely to cause inaccurate representations in the generated images, so we added instructions to the prompt to avoid jargon in the output of (i) (Appendix~\ref{appendix:prompt:image}). We augmented the output of (i) with a styling prompt instructing the text-to-image model to produce an abstract colored image. Specifically, we decided to have our system generate an image without including complex, realistic details by presenting an image in a \textit{cartoon} style. As the model generates a colored image by default, we designated the styling of a colored image by prepending \texttt{"A cartoon of"} to the output prompt from (i). Finally, we generated an image by feeding the complete prompt as input into the text-to-image model.

\subsubsection{Paper summary}

To generate the summary of the paper, we instructed an LLM to summarize the abstract of the paper in two sentences using the following prompt. Here, as the papers are often written in the first person, we engineered the prompt to summarize in the third person (Appendix~\ref{appendix:prompt:summary}).

\subsubsection{Evidence}

The system converts the XML structure into an array that contains every paragraph of the paper, matched to its section title. Then, we prompted an LLM to run a Q\&A task by asking for evidence for the design implication text (Appendix~\ref{appendix:prompt:evidence}), and find out which source paragraph was used to generate the output. We used the LlamaIndex\footnote{https://github.com/run-llama/llama\_index} library to overcome the prompt-length limitation of an LLM and enable in-context learning. Lastly, we programmed the system to use the selected paragraph as evidence for the design implication, appending the section number for reference.

\subsubsection{Citation}

We used the \texttt{doi.org} API to construct a citation for each paper. We designated APA as the citation format in the API request header. For some papers that do not have a DOI, our system includes the paper title as well as the author list as a citation.

\subsection{Implementation}\label{sec:implementation}

Our system is implemented as a web interface, where the user can upload a paper PDF and choose the design implication text from which to generate a design card. While our generated cards are primarily targeted to designers, our interface can be used by any user group---authors, designers, science communicators, and even the general public. The interface is built on a Javascript framework (SvelteKit), which is connected to a Python backend server deployed on AWS EC2 that generates and returns the components of the design card.

More specifically, the server takes as input (i) a PDF of the paper and (ii) a selected design implication text, and outputs a design card in HTML format. First, the system uses the Grobid~\cite{lopez2009grobid} library to parse the paper PDF file into the structured XML format, and the system uses the BeautifulSoup\footnote{https://www.crummy.com/software/BeautifulSoup/} library to process the XML structure and extract components (\ie{}, title, abstract, DOI, paragraphs) used to generate the design cards. The overall pipeline for generating design cards is illustrated in \autoref{fig:implementation}.

For the system implementation tested in our evaluation, we used GPT-3 as our LLM for generating all textual components, and DALL-E 2 for images. These generative AI models are connected to the backend server through an API. For all GPT-3 API requests, we used the \textit{text-davinci-003} model with the following parameters: \texttt{temperature: 0.5, top p: 1, frequency penalty: 0, presence penalty: 0, best of: 1}. We selected these models as they are strong state-of-the-art models for text and text-to-image generation, though we make no claims that they are the optimal models for these tasks. Other models could be easily adapted for use in our modular pipeline.
\section{Evaluation}

To evaluate the feasibility of communicating design implications using design cards generated by our system, we conducted a survey study with designers ($N=21$) and authors of HCI papers ($N=12$).

\subsection{Study Setup \& Procedure}

\subsubsection{Survey study with designers}

In order to understand the perception of designers toward the design cards generated by our system, we set up a within-subjects survey study with designers as follows:

First, we used the same approach as we did in the preliminary study to identify HCI papers in the ACM DL containing design implications. We chose 4 HCI papers with varying themes and research methods for our survey study with designers (\autoref{tab:paper}; paper 1 -- 4). For each paper, we selected a segment of paper text that contains a design implication; if there are multiple implications, we randomly chose one among them. Then, we fed the segment into our system and generated a design card for each paper.

Each participant in our study conducted a survey in our online survey interface. Each participant was asked to evaluate two papers that were randomly chosen, and for each paper, was presented two formats for communicating the design implication: (i) raw text format (a part of the paper that contains a design implication) and (ii) design card format. Here, the only markup presented in the raw text condition was citation markers; a link to the original paper was provided in both conditions to enable participants to reference the original paper when needed. Each evaluation contains six questions for evaluating the perceived qualities of the design implication~\cite{sas2014generating} (\ie{}, empirical/theoretical validity, generalizability, originality, generativity, inspirability, and actionability) on a 7-point Likert scale. To ensure participants’ consistent understanding of the evaluation metrics, our questionnaire incorporated the definitions outlined in the original paper~\cite{sas2014generating} (\eg{}, generativity -- to create and open up new design spaces; p. 1978). In addition, we randomized the order of both papers and formats for every participant to avoid the ordering effect.

Once participants finished evaluating formats for the two papers, they were directed to a screen where they were asked their preferred format for consuming design implications from academic papers. They were also asked to provide the reasons for their preference, as well as suggestions for future enhancements for the design cards.

To recruit designers for our survey study, we posted our recruitment in two design-focused online communities at our university. In this posting, we provided an explanation and goals of our study, along with a link directing to our online survey. As a result, 21 designers completed the survey---13 were working as designers, 7 were pursuing a degree in design, and 1 was unemployed. The average age of the participants was 26.5 ($SD=5.7$); 14 of them were female, 6 male, and 1 preferred not to say. Of all participants, 10 responded that they had heard of the concept of a design card before.

\begin{table}
    \caption{Papers used for generating design cards and running the evaluation}
    \small
    \begin{tabular}{lllll}
    \toprule
    & \textbf{Theme} & \textbf{Venue} & \textbf{\leftcell{Research method}} & \\
    \toprule
    1 & \leftcell{Accessibility,\\game} & CHI '22 & Qualitative & \cite{mason2022including}\\
    \midrule
    2 & \leftcell{Healthcare,\\voice agent} & TiiS, '22 & Mixed-methods & \cite{brewer2022empirical}\\
    \midrule
    3 & \leftcell{Web search,\\UI design} & CHI '20 & Quantitative & \cite{baughan2020keep}\\
    \midrule
    4 & Chatbot & DIS '21 & Quantitative & \cite{svikhnushina2021user}\\
    \midrule
    5 & Healthcare & CHI '20 & Mixed-methods & \cite{kim2020helping}\\
    \midrule
    6 & Online survey & CHI '19 & Quantitative & \cite{august2019pay}\\
    \midrule
    7 & \leftcell{Accessibility,\\interaction technique} & \leftcell{MobileHCI\\'21} & Quantitative & \cite{ahmetovic2021replay}\\
    \midrule
    8 & \leftcell{AI fairness toolkit} & FAccT '22 & Qualitative & \cite{deng2022exploring}\\
    \midrule
    9 & Online community & CSCW '20 & Quantitative & \cite{xia2020exploring}\\
    \midrule
    10 & Authoring tool & UIST '22 & Mixed-methods & \cite{suh2022codetoon}\\
    \midrule
    11 & \leftcell{Social VR,\\online interaction} & CSCW '21 & Qualitative & \cite{freeman2021body}\\
    \midrule
    12 & \leftcell{Accessibility,\\voice-based CA} & CHI '21 & Qualitative & \cite{cha2021exploring}\\
    \bottomrule
    \end{tabular}
    \label{tab:paper}
    \Description{This table includes the list of papers used for our study. 12 papers are listed in this table, and each paper has its corresponding theme, venue, research method, and citation as columns.}
\end{table}

\subsubsection{Survey study with the authors of HCI papers}

To understand if our system can be an effective tool for the paper authors to communicate design implications, we contacted the authors of HCI papers asking them several open-ended questions via email. In addition to the four papers that we used for the survey study with designers, we decided to contact the authors of additional papers from more diverse domains to gain richer and potentially more generalizable insights into their perception of the design cards. In the email, we included a design card that we generated using their paper, along with the corresponding section of the paper used for generating it. Based on these contents, we asked each author to (i) assess the accuracy and clarity of the contents of our design card and (ii) provide any feedback for future improvement to make AI-generated design cards better support communication of design implications. We received responses from the authors of 12 papers in total, including from the authors of all 4 papers that were used for the survey study with designers, as well as 8 other respondents (\autoref{tab:paper}; paper 1 -- 12).

\subsection{Analyses}

\subsubsection{Quantitative analysis}

In our survey study, each participant evaluated the design implications and their corresponding design cards for two (randomly assigned) papers. Thus, to analyze our quantitative results while controlling for the random effect and the paper ID, we used a linear mixed-effects model to test for a significant difference across the perceived qualities of design implications from reading the two formats. Specifically, we modeled this by including the paper ID as a control variable and the participant ID as a random variable.

\begin{figure*}
    \centering
    \includegraphics[width=\textwidth]{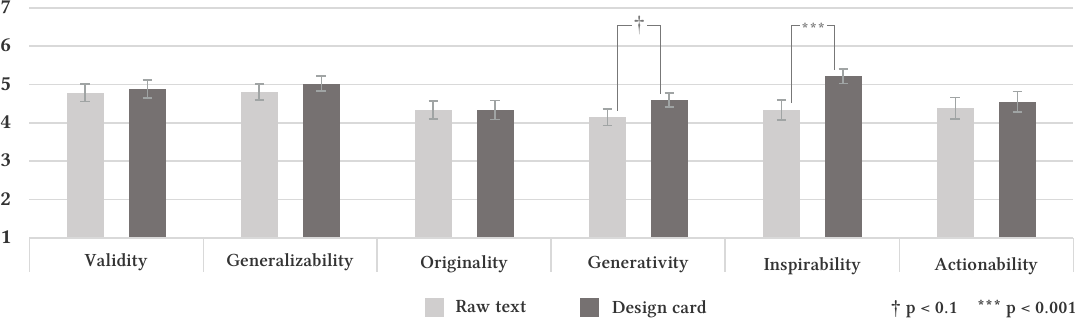}
    \caption{Perceived qualities of design implication for each format using metrics from Sas \etal{} (2014)~\cite{sas2014generating}. Significance levels are based on the generalized linear mixed-effects modeling, and the bars indicate standard errors.}
    \label{fig:survey}
    \Description{This figure illustrates the perceived qualities of a design implication for design cards and raw texts from our study. It contains the evaluation of validity, generalizability, originality, generativity, inspirability, and actionability. Each measure is measured on a 7-point Likert scale. For the validity, generalizability, originality, and actionability, there was no significant difference for both design card and raw text conditions. For generativity and inspirability, the design card condition showed a significantly higher level compared to the raw text condition.}
\end{figure*}

\subsubsection{Qualitative analysis}

Once the studies were complete, we then followed the same approach as we did with the preliminary study data to develop themes. In Subsection~\ref{sec:qual}, we refer to each participant as D1 -- D21 (designers) and A1 -- A12 (authors), with participant order randomized.

\subsection{Quantitative Results}

\subsubsection{Preference on the format}

Our survey revealed that the majority of designers preferred to consume a design implication through design cards. In the survey results, 15 designers responded that they preferred to consume design implications using a design card format, while 3 preferred the raw text format; the other 3 designers had no preference.

\subsubsection{Perceived qualities of design implication}

The linear mixed-effects modeling of our survey results revealed that designers perceived the content from the design card format as more inspiring ($M=5.21$, $SD=1.22$) and generative ($M=4.60$, $SD=1.17$), compared to the raw text format ($M=4.33$, $SD=1.75$; $M=4.14$, $SD=1.42$) (\autoref{fig:survey}). The analysis revealed that there was a significant difference in the perception of inspirability ($t=3.53$, $p<0.001$) and a marginally significant difference in the perception of generativity ($t=1.74$, $p<0.1$).

At the same time, the results showed that design cards were able to deliver a design implication to designers without sacrificing its validity ($M=4.88$, $SD=1.53$ vs. $M=4.79$, $SD=1.51$; $t=0.35$, $p=0.73$), generalizability ($M=5.02$, $SD=1.30$ vs. $M=4.81$, $SD=1.35$; $t=0.81$, $p=0.42$), or originality ($M=4.33$, $SD=1.57$ vs. $M=4.31$, $SD=1.47$; $t=0.09$, $p=0.93$). There was also no significant difference in actionability between design cards and raw texts ($M=4.55$, $SD=1.76$ vs. $M=4.38$, $SD=1.79$; $t=0.48$, $p=0.63$).

\subsection{Qualitative Results}~\label{sec:qual}

\subsubsection{Visual hierarchy and visual elements made design cards appealing and easy to consume}

Participants reported the card format to be well-structured, with a clear visual hierarchy that facilitated information processing. They perceived information to be well-organized, improving the readability of the contents: \textit{“I liked the cards mainly due to their visual hierarchy as I could know what is important and what is not.”}~(D2) Having the key information (\ie{}, title) upfront and background contexts on the back side, design cards allowed participants to quickly gain a sense of a design implication, while allowing them to consume other information if needed: \textit{“I could easily digest by getting an at-a-glance summary of the insight with the title and then clear divisions of a longer explanation of the insight with factual references below.”}~(D18)

In addition, designers from our study evaluated the design of design cards (\eg{}, color, image styling) as visually clear and appealing, which was cited as the reason for their inclination to read the card: \textit{“I loved the colors (of design cards)”}~(D16); \textit{“I enjoyed seeing a visual graphic, which led me to prefer design cards over plain texts.”}~(D11) Moreover, the images of design cards were reported to successfully play a role in assisting the designers to easily grasp the whole context of a design implication. Designers mentioned that the context depicted in an image was catchy, enabling them to easily consume the design implication: \textit{“The art on the cards successfully served its purpose of illustrating what information was being shared (in the paper).”}~(D14) Authors also highlighted the image's potential use of assisting readers in comprehending the connection between the target user and the design context of the paper: \textit{“The picture describes the relationship and interaction between [target user of the paper] and [design artifact] well. The most eye-catching part.”}~(A9)

\subsubsection{Our design cards afforded digestibility and ease of understanding through concise language}

The raw text format was perceived as more difficult to read. With a long and dense text block, raw text format was reported to make it harder for participants to follow the contents, causing them to lose interest easily and become bored: \textit{“I found it harder to absorb a large bit of text from the raw text.”}~(D14); \textit{“I felt the raw text format is too boring to read.”}~(D7) In contrast, participants found the card format easy to understand and digest compared to the raw text format. They appreciated the use of concise language enabled by automated summarization, which helped them easily identify key points and main takeaways: \textit{“Design insights were given to me without me having to read too much or decipher from the source text.”}~(D21)

Aligning with the perception of designers, the authors of the paper who responded to our questions similarly reported the clarity and effectiveness of design cards to be satisfactory. They also expressed satisfaction with the idea of automatically generating design cards, noting that these cards accurately reflect their design implication well: \textit{“(The design card) correctly reflects what we stated in [section] and the wording of the [description] is relatively similar to the text in the paper - it's good.”}~(A8); \textit{“The idea is very cool and I think your system does a good job in summarizing the design implication from [section] in a crisp and concise way.”} (A6) With such well-aligning contents presented in a convenient format, authors viewed design cards as an effective means of communicating their design implications: \textit{“The generated design card communicates the design implication listed in [section] of our paper effectively.”}~(A2). At the same time, authors did suggest a few potential refinements of generated texts in our design cards, such as reducing factual misalignment from the paper summary when combining multiple research findings simultaneously to achieve brevity~(A6) and enhancing descriptions to more accurately reflect the role of the design target~(A5).

Based on the clarity and effectiveness of the format, the authors highlighted the potential of design cards for strengthening their contributions around the design implications. By collaborating with our design card system, authors believed that our system could be a great authoring tool for augmenting the delivery of their paper contributions: \textit{“If researchers were to go through one or two revisions based on this approach, it could indeed greatly enhance the delivery of their contribution much strongly and in a more meaningful way.”}~(A3)

\subsubsection{Our system supported communication of design implications without loss of validity and originality}

From our quantitative analysis, we identified that design cards delivered a design implication without sacrificing any validity and originality. Aligning with these results, design card and raw text formats were both reported to be valid and believable, even if the design card was programmatically generated from the original paper text. In particular, we identified that the inclusion of extracted evidence and the paper citation contributed to making the design card a reliable source for design insights. Such cues implying that the results were verified led participants to perceive the design card as more legitimate: \textit{“Seeing the parts on the bottom (evidence, citation) made it look professional.”}~(D19)

At the same time, the survey with the authors revealed the need to refine the pipeline for extracting evidence for a design implication. To identify evidence from the paper, we prompted an LLM to run a Q\&A task and programmed our system to use the paragraph of a paper that was used to derive the answer, which was reported to be an effective way to deliver the author's reasoning of the design guideline: \textit{“I like (evidence)... It helps effectively communicate the rationale for the design implications.”}~(A11) Yet, four authors mentioned that the evidence included in the design card of their paper could be potentially enhanced, by pointing out that it was partially correct, not the only evidence present in the paper, or misleading: \textit{“The part about where the guideline came from seems off.”}~(A5) As a potential mitigation strategy, one author offered that having a rule-based benchmarking tool to verify how much the evidence overlaps with the design implication might help: \textit{“Some checks based on word overlap between the generated guideline and the text in the referenced paper paragraph could be helpful to get a proxy for the accuracy.”}~(A2)

\subsubsection{Opportunities for enhancing the use of images}

Our analysis revealed that both designers and authors perceived the use of auto-generated images to help support the communication of design implications: \textit{“The approach of using an AI-generated image for authoring the design implication is interesting and will be super helpful (when communicating design implications).”}~(A3) At the same time, our interview also revealed some potential uses for additional images in a design card to help provide more information about the paper or exemplify the design implication. For example, designers mentioned that adding infographics depicting results from the paper would further help them to understand the study: \textit{“I also want to see a pure infographic for the results.”}~(D13) One author also suggested using a figure directly from their original papers to help illustrate the design guideline as was envisioned: \textit{“I think showing a representative figure of the paper can be helpful, especially if that figure/artifact exemplifies the design mentioned in the design implications or can further clarify.”}~(A11) Another author mentioned that there are multiple key messages they intended to communicate in their design implications section, but that the single image shown reflected only one of them: \textit{“I feel that the image only covers a subset of the design implication presented in the text and may not necessarily highlight all the key messages.”}~(A4) This suggests iterating on our card designs to support the inclusion of multiple images per design card.

\begin{figure}[t!]
    \begin{subfigure}{\linewidth}
        \centering
        \includegraphics[width=\linewidth]{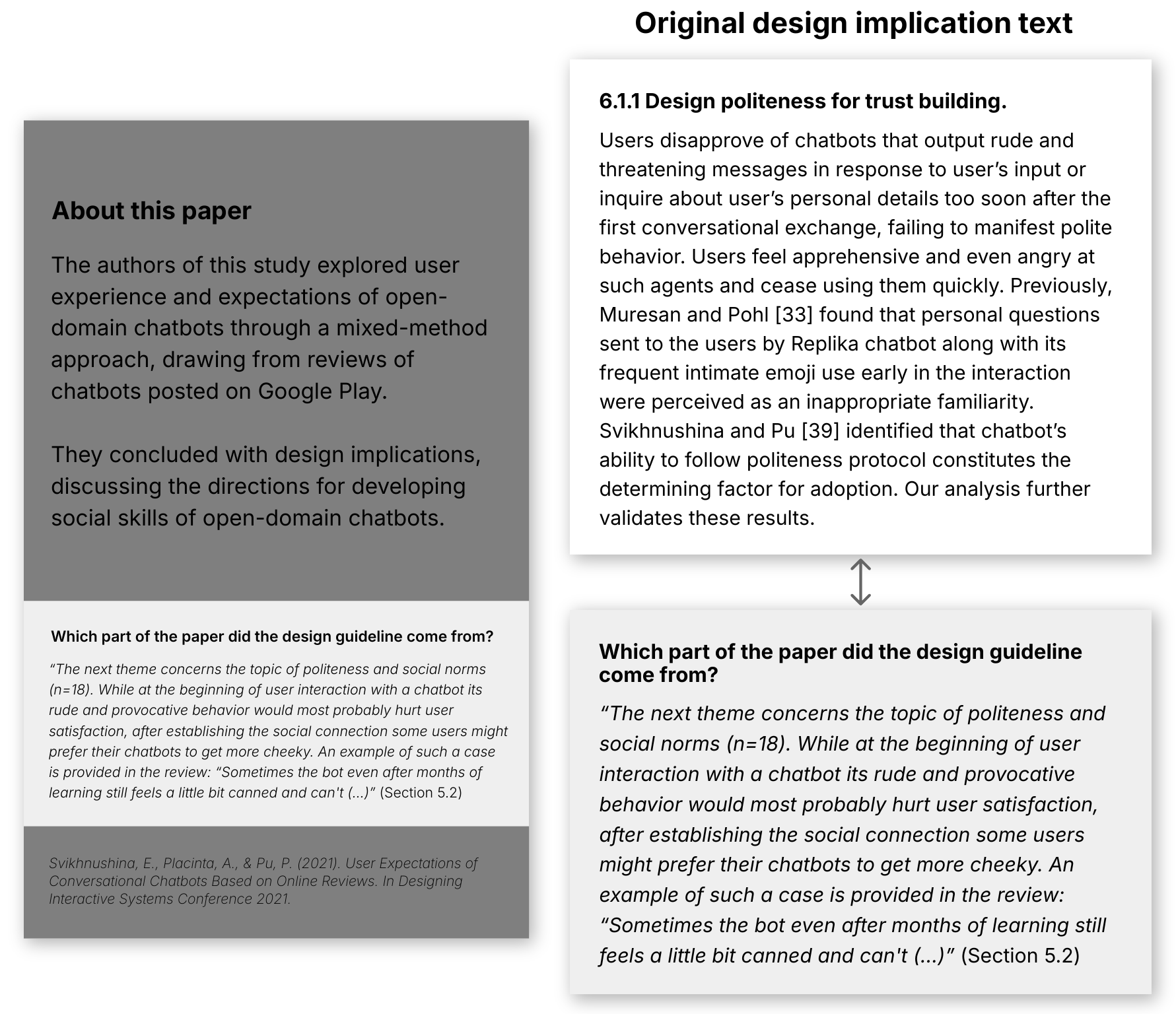}
        \caption{Imperfect match between content and evidence}
        \label{fig:card:error1}
    \end{subfigure}%
    \vspace{.4cm}
    \begin{subfigure}{.47\linewidth}
        \centering
        \includegraphics[width=\linewidth]{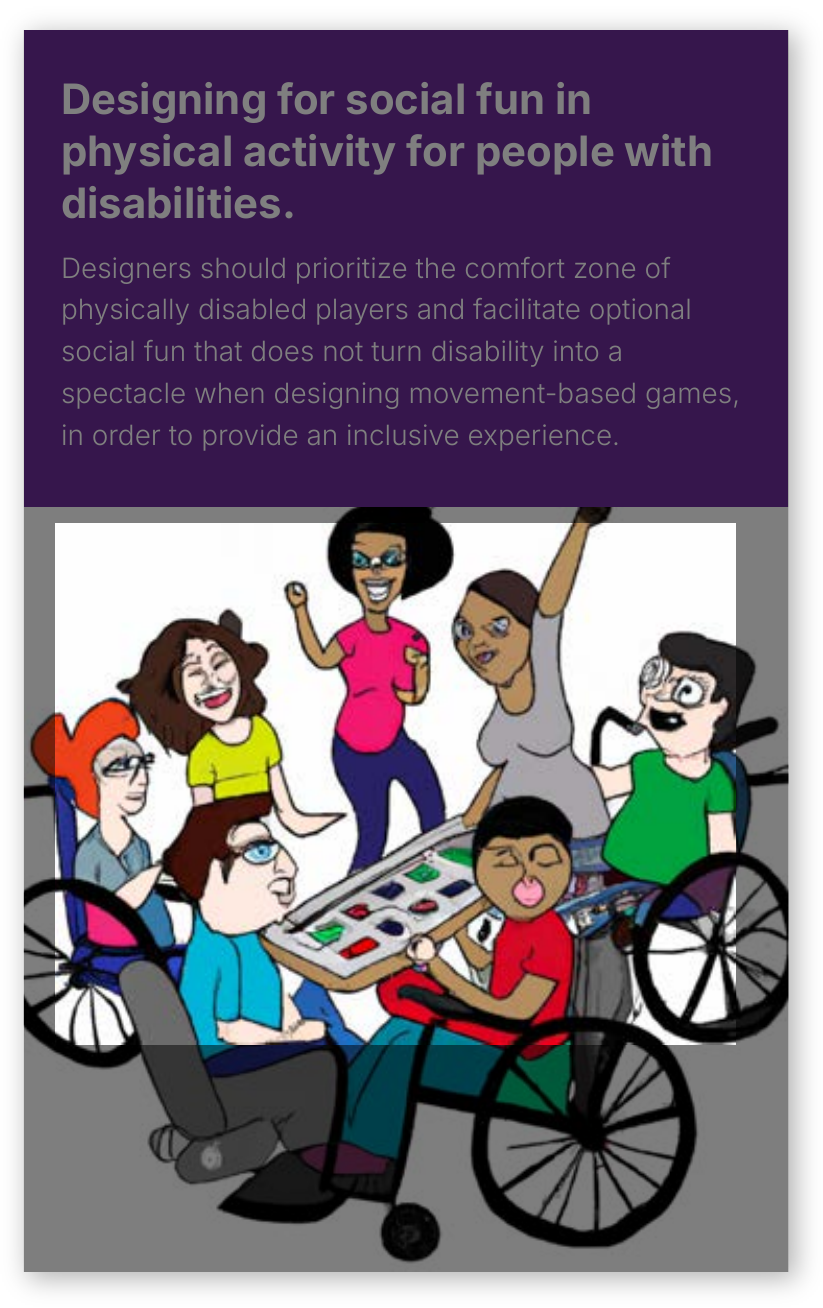}
        \caption{Improper representation of the target users}
        \label{fig:card:error2}
    \end{subfigure}%
    \hspace{.05\linewidth}
    \begin{subfigure}{.47\linewidth}
        \centering
        \includegraphics[width=\linewidth]{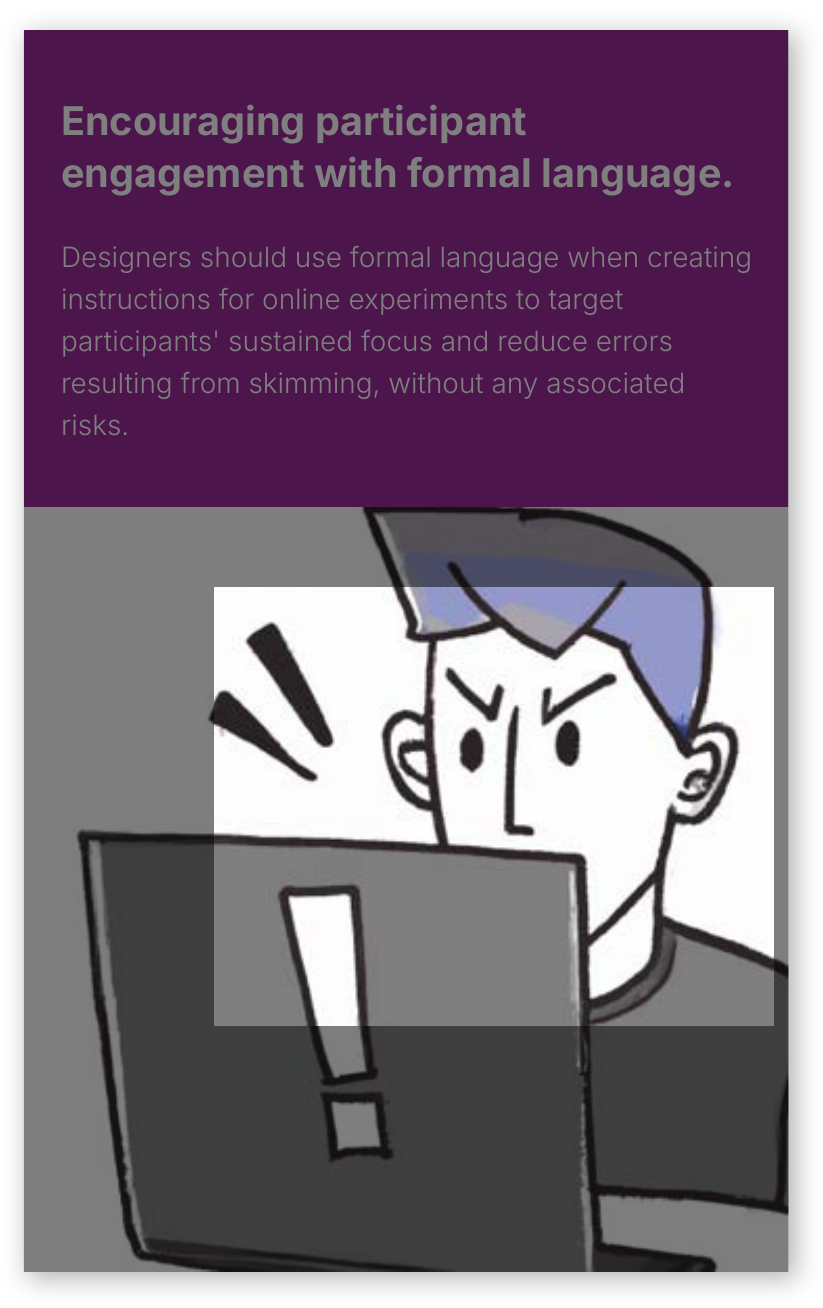}
        \caption{Unintended emotion of the target user}
        \label{fig:card:error3}
    \end{subfigure}%
    \caption{Examples of misalignments surfaced in our survey}
    \label{fig:error}
    \Description{These figures illustrate three of the misalignments that were identified from our survey. The first subfigure illustrates an imperfect matching of the evidence of our design cards, the second subfigure illustrates a misrepresentation of the target users in an image, and the third subfigure illustrates an unintended emotion of the target user in an image.}
\end{figure}

Despite the generally positive perceptions towards the generated images, a few authors did note some misalignment between their expectations toward the image representation and the actual image (\eg{}, \autoref{fig:error}-b, c). For example, there was a misrepresentation of the target users in an image that depicts people with disabilities, which the author found to be inappropriate~(A12). Other authors also noted a misalignment between their intent and the output. For example, A1 pointed out the misalignment between their expected emotion toward potential users and that of the user illustrated in the design card image. Similarly, after seeing a card image featuring two users where only one was using a computing device, A10 clarified that the intended representation should depict both users engaging with the device to ensure clear communication of their design implication's key message of device-mediated communication and avoid any potential confusion.
\section{Discussion}

In this study, we examined the feasibility of AI-generated design cards as a tool for communicating design insights between paper authors and designers. Our system automatically produces design cards based on a design implication from a paper, and demonstrated its potential to assist authors in communicating their design implications. We expect that a more effective process for presenting design implications could help alleviate the burden on paper authors to communicate their findings to design practitioners. Bridging this translational gap can further enhance the impact of valuable design insights from academic papers.

Results from our evaluation suggest that designers perceived our AI-generated design cards as more inspiring and generative compared to traditional paper formats. We also found that these gains were had without sacrificing the validity, originality, and generalizability of the design finding. A key reason for this that we identified in our qualitative analysis is that our system lets designers view the original text via evidence quotes and citations, which ultimately assured them that design card contents were legitimate. In addition, we believe that our design choice of abstracting and stimulating the visual element (\ie{}, abstract, colored representation of the image)---factors known to inspire designers as suggested by Sas \etal{}~\cite{sas2014generating}---has also contributed to such results. However, our survey with paper authors revealed that extracted evidence for a design implication can sometimes be inaccurate or not perfectly aligned with the part of the paper authors used to derive the design implication. As such a misalignment might affect the overall credibility of generated design cards, future systems should focus on validating the accuracy of represented information. Involving authors in the process of generating design cards could help guarantee the validity and appropriateness of outputs.

Our results also suggest that there are opportunities to involve the paper authors in enhancing the use of images on the design cards. Despite the efficacy of AI-generated images in providing an overall context of the design implication, a few authors from our study reported a misalignment between the card images and their expectations based on their original intent (\eg{}, unintended emotion of the depicted user), or expressed a wish to include a representative figure directly from the paper to further help readers' understanding. To address this, we believe that including authors in the loop may help to better reflect their original intent; for instance, the system may offer multiple image prompt candidates, and let authors choose the one that best aligns with their intended message. Likewise, the system may complement our AI-generated image by detecting several representative figures from the paper and letting authors choose and add one that best exemplifies their intended message. As such, we believe that our system could further ensure the alignment between images and the author's intent, without requiring significant efforts.

Despite the effectiveness of our system, one remaining question is how to make the content of design cards generated by our system more actionable. Although we designed our title to involve action, our survey with designers revealed no significant difference in actionability between the design card format and raw text format. One possible reason is that the papers' design implication sections (our study comparisons) were already well-written and were calling for designers to take a particular design action. Thus, designers did not perceive a difference in actionability. However, our study does suggest some changes that we can explore to make the design card even more actionable, such as having an additional component that gives more detailed guidance (\eg{}, step-by-step guidance for designing) or examples of connecting the design implication to real-world artifacts (\eg{}, how a well-known design was/can be enhanced by applying this implication).


Regarding the target audience for our system and the generated artifact, we intended the generated design cards to be specifically read and consumed by designers. Meanwhile, the use of our interface for generating cards is more versatile (Section~\ref{sec:implementation}); we anticipate that various individuals, such as paper authors, designers, science communicators, or the general public, could find value in creating cards using our interface. For example, we envision that authors of newly published papers might utilize our interface to generate design cards to showcase their work to designers. Also, designers may directly use our system to create cards in advance of their actual design activities, enabling the use of these cards for purposes where reading the papers may not be feasible (\eg{}, conducting a design sprint or workshop under time constraints). Investigating how different stakeholders may find value in using our interface to generate cards is another avenue of research.

In addition to understanding such varying purposes of using our interface, another potential consideration that will need to be studied is how to make our card-based translational resource available to designers and improve their awareness. While our system has the potential to improve designers' use of the abundance of rich insights in academic papers, recent research suggests that designers often lack awareness of the wealth of design cards that exist and their potential benefits, despite a set of existing design cards that can be readily used~\cite{roy2019card, hsieh2023designcards}. This awareness problem could worsen if our system helps generate a large influx of cards. One possible solution is to integrate our system into digital libraries (\eg{}, ACM Digital Library) and present the generated design cards alongside each paper. Other more designer-friendly repositories should also be explored to help the dissemination of this type of resource.
\section{Limitation \& Future Work}

There were several assumptions that we made in building and evaluating our system. First, although we evaluated the qualities of design cards generated by our system, we did not test how generated design cards would be used in the real-world designing contexts of designers. From the findings of our controlled study design which may serve as a crucial first step for examining such real-world uses, we should further explore how AI-generated design cards affect the design process and outputs of designers. Future work may also investigate further customizability and generalizability of our concept by (i) exploring how to tailor the use of design cards for different purposes and designer subgroups and (ii) evaluating more papers beyond what we used in our controlled study.

Our interface also assumes a context where the paper segments containing design implications need to be manually localized. While design implication sections are frequently included in HCI publications, future work could investigate ways to automatically identify text corresponding to design implications, as well as segment these into multiple distinct implications. These steps are needed to further streamline the generation of design cards from HCI papers.

Additionally, although our focus was to demonstrate the potential of AI-generated cards by choosing reasonable methods, there are many alternate models, techniques, and prompts that could be used to build our system. Since our system is modular, each component can easily accommodate alternate techniques or models to test performance, and we believe this can be expanded on in future work. Similarly, although our characterization of design card components was based on our cursory examination of existing design card sets, there may also exist alternative design components to achieve our goal of creating useful translational resources for designers (\eg{}, including different card elements for inspirational purposes). Investigating other component options could make these resources richer and allow us to understand the perception of different types of components per user group.

Lastly, future work should strive to mitigate AI biases. As utilizing AI-generated images may retain biases from the datasets they are trained on, it is an open concern~\cite{mishkin2022risks, brown2020language} that the outputs manifest gender and racial stereotypes (\eg{}, higher likelihood of featuring white men rather than representative users; negative associations for certain racial/ethnic groups). To overcome this, it is necessary to develop methods to identify and mitigate such biases in future systems.
\section{Conclusion}

In this paper, we propose an approach of using generative AI models to help create design cards that distill academic findings into a more accessible and prescriptive format. Based on the preliminary interview sessions with design practitioners, we developed a system that automatically generates a design card using generative AI models. Results from our evaluation with designers and HCI paper authors demonstrated the potential of AI-generated design cards in communicating design implications, as well as their future enhancements.

\begin{acks}
This work was supported by an Accelerating Foundation Models Research Award from Microsoft.
\end{acks}

\bibliographystyle{ACM-Reference-Format}
\bibliography{99-bibliography}

\appendix
\section{Study Detail of the Preliminary Study}
\vspace{-.1cm}
\begin{table}[H]
\caption{Participants' demographic and background information of our preliminary study}
\small
\begin{tabular}{lllll}
\toprule
\textbf{ID} & \textbf{Age} & \textbf{Gender} & \textbf{Design focus / background} & \textbf{\leftcell{Design\\work\\experience\\(years)}}\\
\toprule
P1 & 30 & Female & \leftcell{Masters student in\\professional design program,\\user interface/experience\\designer} & 3 - 4 \\
\midrule
P2 & 21 & \leftcell{Non-\\binary} & \leftcell{Urban designer, previous\\experience in urban design\\and CS} & 1 - 2 \\
\midrule
P3 & 24 & Female & \leftcell{User experience designer} & 3 - 4 \\
\midrule
P4 & 22 & Male & \leftcell{Design researcher, user\\interface/experience designer} & 3 - 4 \\
\midrule
P5 & 24 & \leftcell{Non-\\binary} & \leftcell{Masters student in design,\\previously worked as a visual\\designer, BFA in design} & > 5 \\
\midrule
P6 & 25 & Female & \leftcell{Masters student in\\professional design program} & - \\
\midrule
P7 & 24 & Female & \leftcell{Masters student in\\professional design program} & 3 - 4 \\
\midrule
P8 & 25 & Female & \leftcell{Design researcher} & 2 - 3 \\
\midrule
P9 & 28 & Female & \leftcell{User experience designer} & 3 - 4 \\
\midrule
P10 & 24 & Female & \leftcell{Masters student in\\professional design program,\\previous experience in\\software engineering} & - \\
\midrule
P11 & 23 & Female & \leftcell{User experience designer \&\\design researcher, focusing\\on accessible designing} & 1 - 2 \\
\midrule
P12 & 33 & Female & \leftcell{Product designer} & 4 - 5 \\
\bottomrule
\end{tabular}
\label{tab:participants}
\Description{This table lists the demographic and background information of the designers who participated in our preliminary study. There are 12 rows in this table, each of which details the participant's information, and each row has ID, age, gender, design focus/background, and design work experience about each participant as columns.}
\end{table}

\section{Prompts Used for Generating Card Components}

\subsection{Title}\label{appendix:prompt:title}

\texttt{The following text snippet is part of an academic paper (title: \{PAPER TITLE\}), which contains a guideline for a certain design practice. From this text, extract a concise title that delivers what action designers should take in gerund-format (\ie{}, verb+ing): \{DESIGN IMPLICATION TEXT\}}

\subsection{Description}

\subsubsection{Generating description text}\label{appendix:prompt:description:text}

\texttt{The following text snippet is part of an academic paper (title: \{PAPER TITLE\}), which contains a guideline for a certain design practice. From this text, extract a one-line argument for designers that contains (i) who the design is targeting, (ii) what the design subject is, and (iii) the justification for the design: \{DESIGN IMPLICATION TEXT\}}

\subsubsection{Verifying description text}\label{appendix:prompt:description:verification}

\texttt{Does the following statement contain the \{ENTITY (e.g. justification)\} for designers to follow a certain design direction? Answer in YES or NO: \{GENERATED DESCRIPTION TEXT\}}

\subsection{Image}\label{appendix:prompt:image}

\texttt{Illustrate a one-line usage context for the following design guideline. The output should not contain any jargon:\\\\Design guideline: Since older adults are unfamiliar with video medium, (...)\\Usage context: an elderly person looking at a laptop with a video playing on it\\..\\\{FEW-SHOT EXAMPLES\}\\..\\Design guideline: \{DESIGN IMPLICATION TEXT\}}

\subsection{Paper Summary}\label{appendix:prompt:summary}

\texttt{Summarize the following text in two sentences, describing the authors in the third person: \{ABSTRACT\}}

\subsection{Evidence}\label{appendix:prompt:evidence}

\texttt{What can be the evidence for this design implication: \{DESIGN IMPLICATION TEXT\}\\(with in-context learning of the indexed paper paragraphs)}

\section{Examples of Design Cards Generated from Our System}\label{appendix:card}

\begin{figure*}[h!]
  \centering
  \begin{subfigure}{.45\textwidth}
    \centering
    \includegraphics[width=\linewidth]{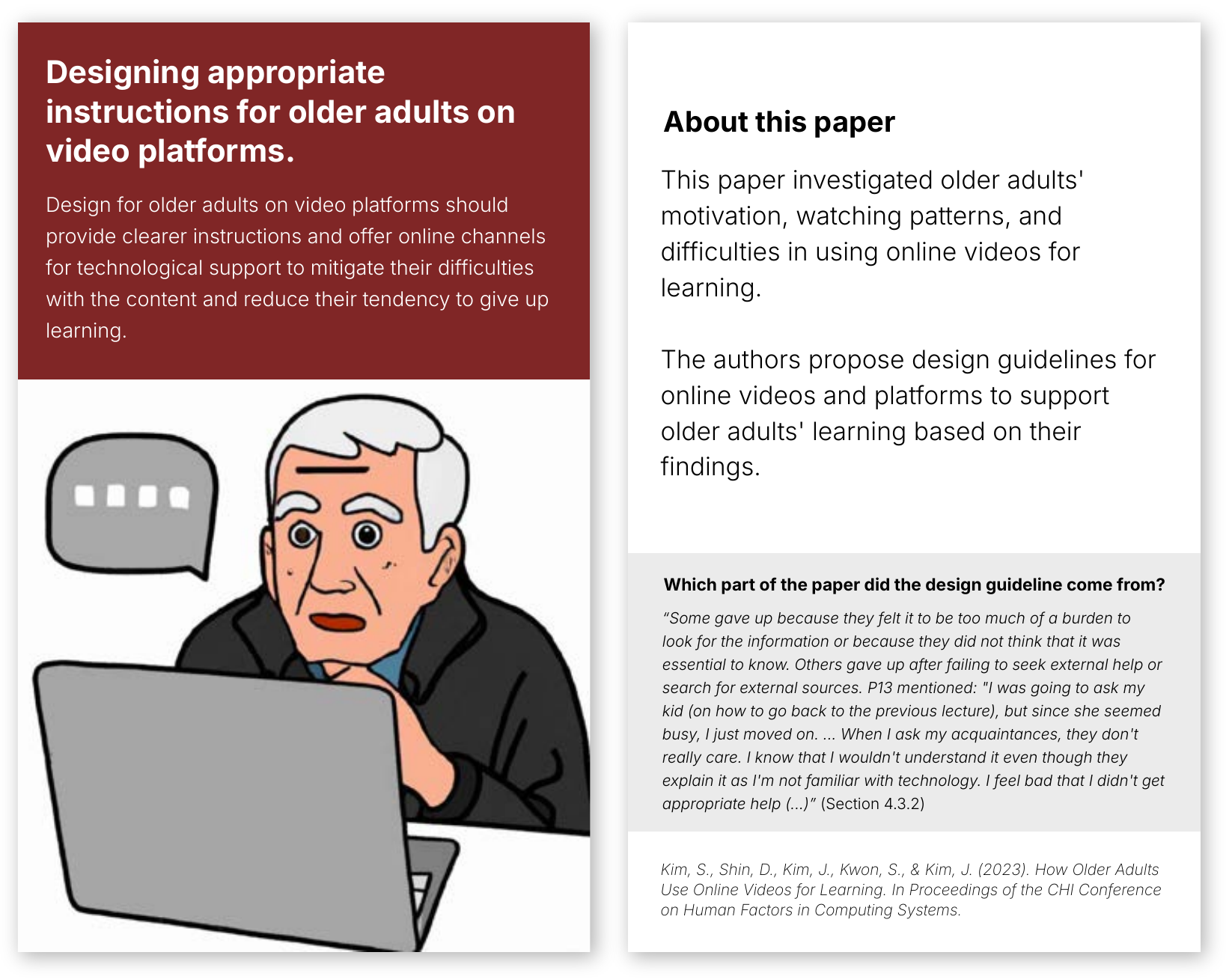}
        \caption{Kim \etal{}, 2023~\cite{kim2023older}}
        \label{fig:card:example1}
  \end{subfigure}
  \vspace{.5cm}
  \begin{subfigure}{.45\textwidth}
    \centering
    \includegraphics[width=\linewidth]{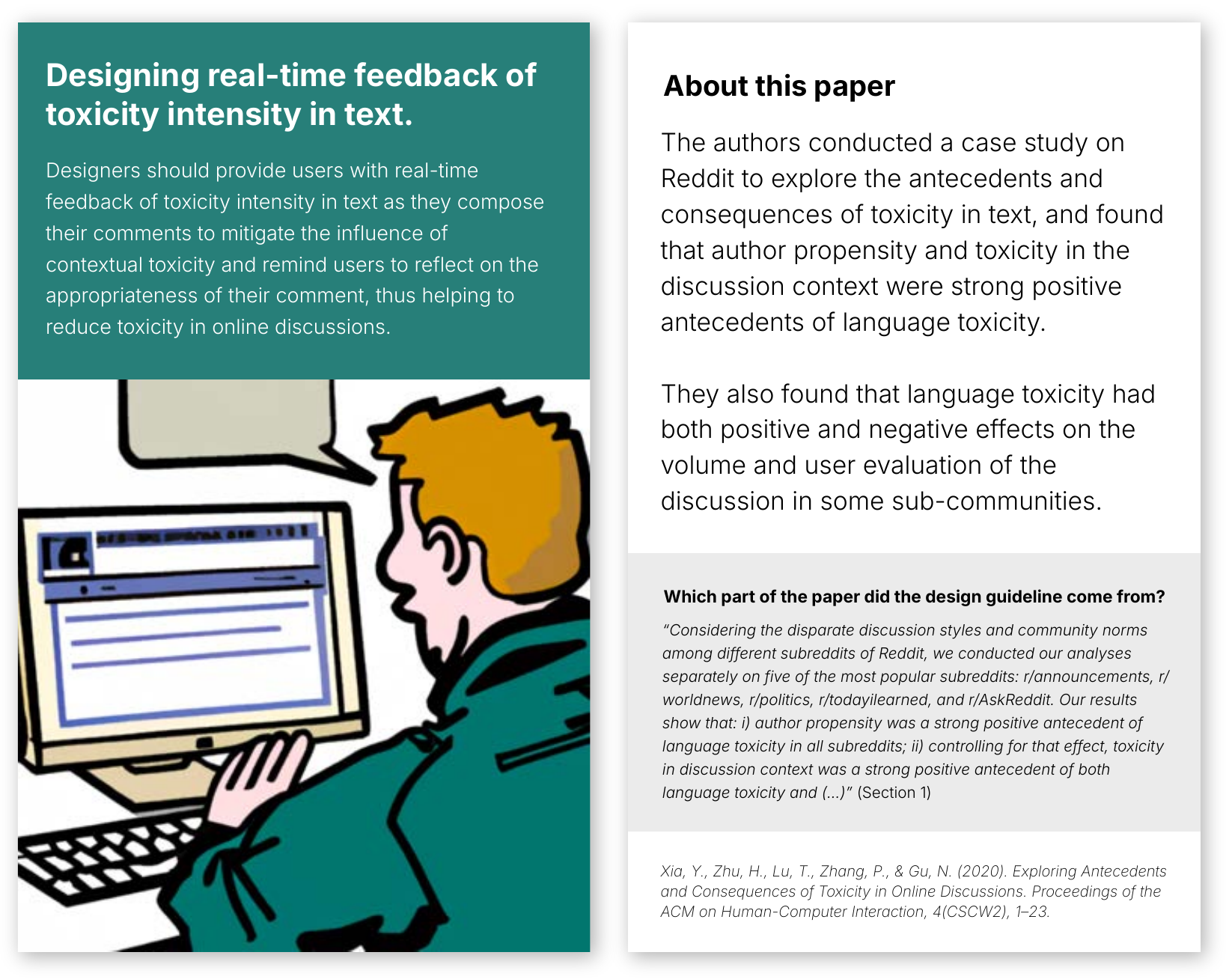}
        \caption{Xia \etal{}, 2020~\cite{xia2020exploring}}
        \label{fig:card:example2}
  \end{subfigure} 
  \begin{subfigure}{.45\textwidth}
    \centering
    \includegraphics[width=\linewidth]{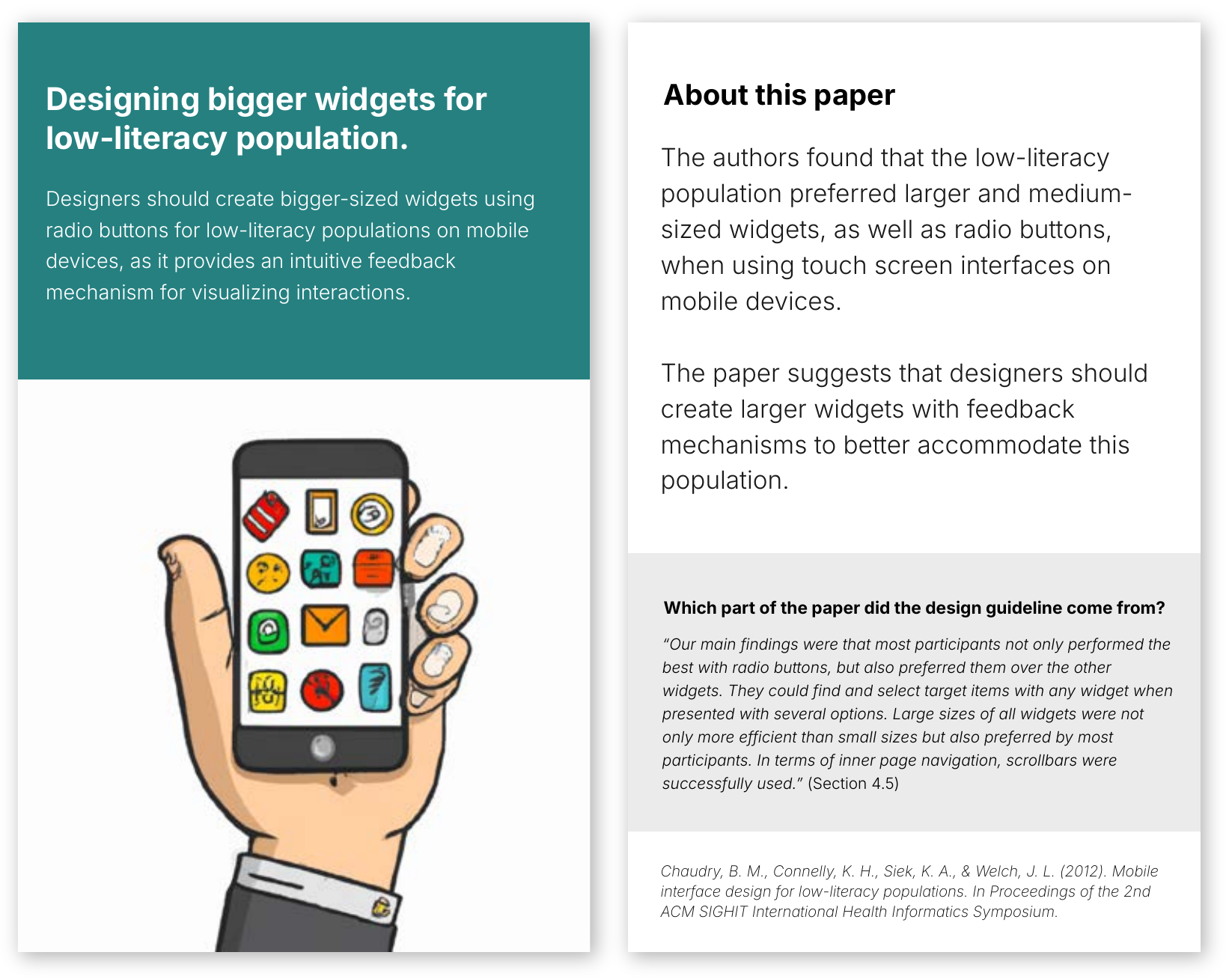}
        \caption{Chaudry \etal{}, 2012~\cite{chaudry2012mobile}}
        \label{fig:card:example3}
  \end{subfigure}%
  \begin{subfigure}{.45\textwidth}
      \centering
      \includegraphics[width=\linewidth]{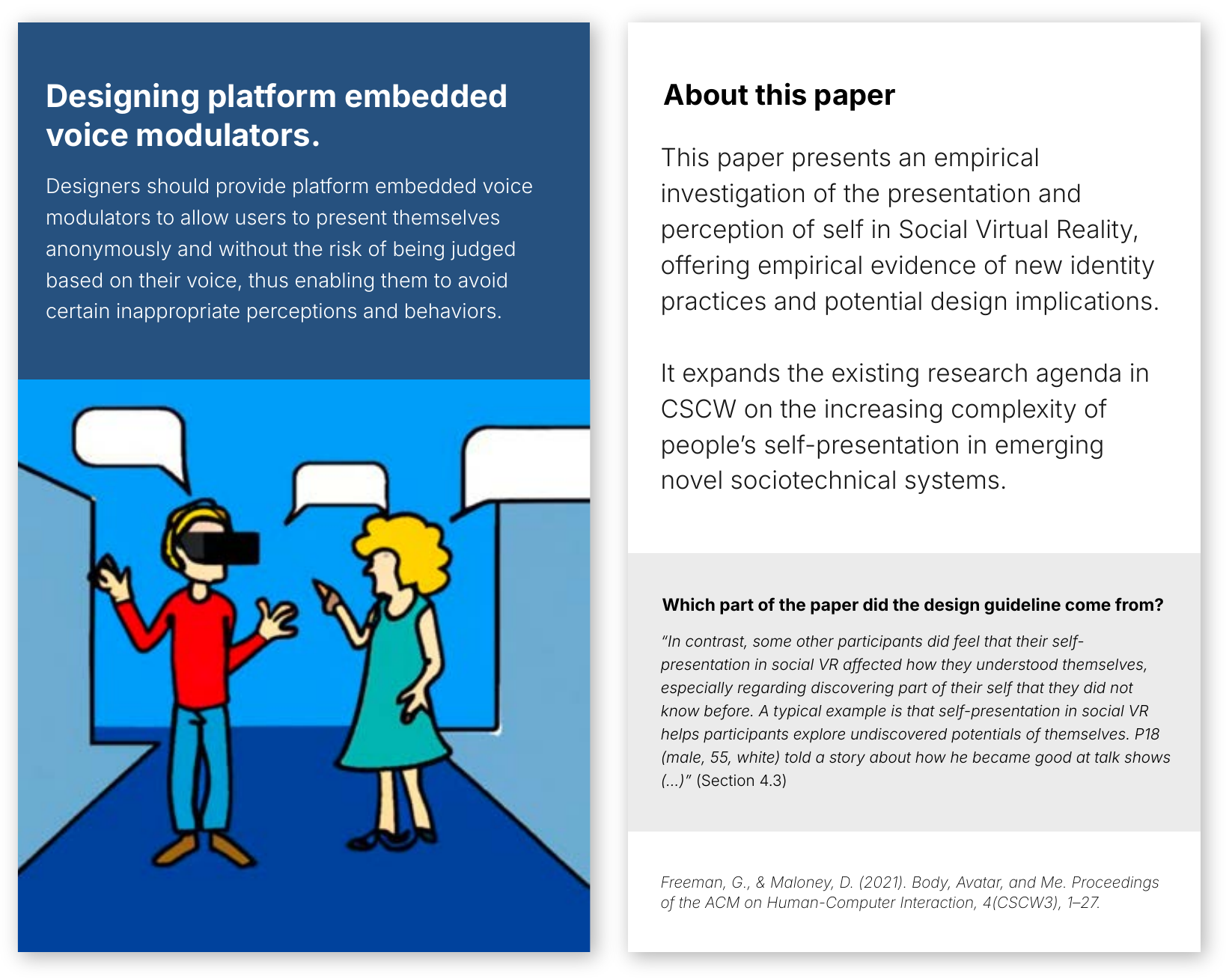}
        \caption{Freeman \& Maloney, 2021~\cite{freeman2021body}}
        \label{fig:card:example4}
  \end{subfigure}%
\caption{Examples of design cards generated from our system}
\label{fig:card}
\Description{These figures illustrate four sample design cards generated from our system. The first design card is generated from a paper about older adults' use of online videos for learning, and the second design card is generated from a paper that explores toxicity in online discussion. The third design card is generated from a paper that speaks to the importance of designing larger widgets for aiding the low-literacy population in using mobile devices, and the fourth design card is built on a paper that explores the presentation and the perception of self in social VR.}
\end{figure*}

\end{document}